\documentclass[preprint2]{aastex}
\usepackage{amsmath}
\newcommand{\sout}[1]{}   
\newcommand{\vect}[1]{\mbox{\boldmath $#1$}}
\shorttitle{Simulations of gyrosynchrotron emission}
\shortauthors{Kuznetsov et al.}
\begin{document}
\title{3D simulations of gyrosynchrotron emission from mildly anisotropic nonuniform electron distributions in symmetric magnetic loops}
\author{Alexey A. Kuznetsov\altaffilmark{1,2}, Gelu M. Nita\altaffilmark{3}, \& Gregory D. Fleishman\altaffilmark{3,4}}
\altaffiltext{1}{Armagh Observatory, Armagh BT61 9DG, Northern Ireland}
\altaffiltext{2}{Institute of Solar-Terrestrial Physics, Irkutsk 664033, Russia}
\altaffiltext{3}{Center For Solar-Terrestrial Research, New Jersey Institute of Technology, Newark, NJ 07102}
\altaffiltext{4}{Ioffe Institute, St. Petersburg 194021, Russia}
\begin{abstract}
Microwave emission of solar flares is formed primarily by incoherent gyrosynchrotron radiation generated by accelerated electrons in coronal magnetic loops. The resulting emission depends on many factors, including pitch-angle distribution of the emitting electrons and the source geometry. In this work, we perform systematic simulations of solar microwave emission using recently developed tools (GS Simulator and fast gyrosynchrotron codes) capable of simulating maps of radio brightness and polarization as well as spatially resolved emission spectra. A 3D model of a symmetric dipole magnetic loop is used. We compare the emission from isotropic and anisotropic (of loss-cone type) electron distributions. We also investigate effects caused by inhomogeneous distribution of the emitting particles along the loop. It is found that  effect of the adopted moderate electron anisotropy is the most pronounced near the footpoints and it also depends strongly on the loop orientation. Concentration of the emitting particles at the loop top results in a corresponding spatial shift of the radio brightness peak, thus reducing  effects of the anisotropy. The high-frequency ($\gtrsim 50$ GHz) emission spectral index is specified mainly by the energy spectrum of the emitting electrons; however, at  intermediate frequencies (around 10-20 GHz), the spectrum shape is strongly dependent on the electron anisotropy, spatial distribution, and magnetic field nonuniformity. The implications of the obtained results for the diagnostics of the energetic electrons in solar flares are discussed.
\end{abstract}

\section{Introduction}\label{S_Intr}
Microwave emission produced during solar flares is known to contain highly important  information about fast electron acceleration and transport, coronal magnetic field and thermal plasma \citep{Bastian_etal_1998, Fl_etal_2009}. However, this potential of the microwave emission has not yet been converted to a routine diagnostics because of two main reasons. The first of them is absence of well-calibrated radio imaging spectroscopy data with needed spatial, temporal, and spectral resolutions \citep{Gary_2003, gary_keller_2004}. Currently, the situation has started to change as a number of solar radio instruments (e.g., Owens Valley Solar Array, OVSA, and Siberian Solar Radio Telescope, SSRT) experience significant upgrade (Upgraded SSRT, USSRT) and expansion (Expanded OVSA, EOVSA), while even more powerful solar radio facilities are planned to be built in near future, and a general-purpose radio facility, EVLA, will soon be operational in the solar observing mode. This implies that the required radio data will soon become available. However, even with them there is a second reason, which is apparent lack of realistic 3D modeling of the microwave emission from flares. This modeling is highly important because the gyrosynchrotron (GS) emission depends in a complicated nonlinear way on many involved parameters and source geometry including spatial inhomogeneity and angular anisotropy. The  realistic modeling has to establish a clear quantitative picture and solid detailed understanding of how the involved physics (i.e., source properties and parameter regimes) affects the emission produced, e.g., what changes in the emission can be expected from variation of a given parameter.

Available 3D models of the GS emission \citep{Preka_Alis_1992, Kucera_etal_1993, Bastian_etal_1998,
Tzatzakis_etal_2008, Simoes_Costa_2006, Fl_etal_2009, Simoes_Costa_2010} have established valuable examples of the flaring microwave emission; however, they are neither numerous nor comprehensive and rely on isotropic pitch-angle distribution and uniform spatial distribution in most of the cases. The observations, however, suggest that fast electrons in flares have often anisotropic and/or inhomogeneous distributions \citep[e.g.,][]{melnikov_etal_2002, spikes, Fl_2005n, Altyntsev_etal_2008}.

Analysis of the pitch-angle anisotropy effect  has yielded controversial conclusions: although the anisotropy was found to have huge effect on the GS emission from  a uniform (or spatially resolved) source \citep{FlMel_2003b, spikes}, this effect can become much weaker when averaging over significant volume with a nonuniform loop magnetic field comes into play \citep{Simoes_Costa_2010}. This calls for a more systematic study of the GS emission from anisotropic electron distributions in 3D magnetic loops. The situation is additionally complicated by the fact that the mentioned pitch-angle anisotropy implies, as a side effect, a spatial inhomogeneity of the electron distribution due to the fast electron accumulation at the loop-top \citep{Lee_Gary_Zirin_1994, melnikov_etal_2002}.

As a first step towards addressing the whole problem of GS modeling in realistic 3D coronal magnetic loops, this paper presents a convenient modeling tool, ``3D GS Simulator'', which gives any user an ability to build an analytical dipole magnetic loop, select a desired viewing angle, populate the loop with thermal plasma and nonthermal electrons, and compute microwave images and spatially resolved spectra. The vacuum-like propagation from the source to the observer is adopted in the tool, so any propagation effect on the radiation in the coronal plasma including polarization  modification, which might be present in the reality, is ignored. A reasonably quick computation of a given model is made possible by the use of recently developed fast GS codes \citep{Fl_Kuzn_2010} that proved to be highly accurate for both isotropic and anisotropic electron distributions. We intentionally consider here analytical models for the magnetic field and electron distribution to fully control the input and reliably interpret the outcome; numerical input in form of corresponding datacubes is under development and will be presented soon elsewhere. This paper considers the microwave (GS + free-free) emission produced by mildly anisotropic electron distributions from a symmetric dipole magnetic loop viewed at different angles and for different parameter combinations. We find that even in such non-extreme conditions the anisotropy  makes noticeable fingerprints on the emission properties, which are discussed in detail below.

\section{Simulation tool and method}\label{S_Simulations}
As has been said, the diversity of the microwave flaring emission is huge, so neither a paper nor a paper series can, perhaps, offer a truly comprehensive table of options fully covering all relevant parameter combinations. Thus, entirely different approach to address the whole problem of the GS modeling is called for: we need a simulation tool capable to smoothly change all the involved source parameters and quickly compute and return the datacubes describing the microwave emission produced. Here, we present such a tool and give an example of using it for the microwave emission simulation.


\subsection{Dipole magnetic flux tube model}\label{tool}
\begin{figure}[h]
\centerline{
(a)\resizebox{4cm}{!}{\includegraphics[clip=true]{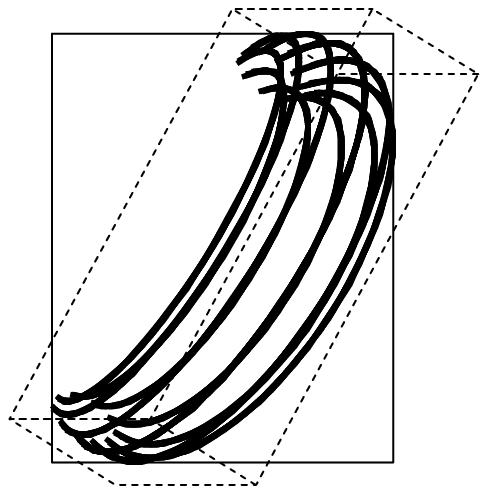}}
(b)\resizebox{4cm}{!}{\includegraphics[clip=true]{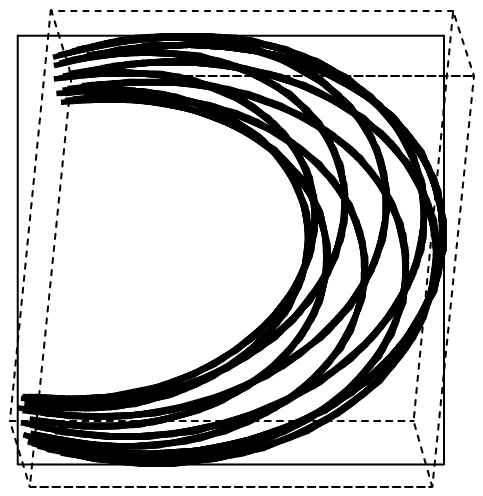}}
}
\caption{The magnetic field model used in the simulations, as implemented in the GS Simulator tool. The dashed box inscribes the portion of the magnetic loop (visualized by a few bold field lines) situated above the solar surface, while the solid rectangle represents the top view of an inscribing box that is perpendicular to the observer's line of sight. The two panels show two different orientations of the same model corresponding to a loop located near the center of the solar disk (a) and a loop located near the solar limb (b).}
\label{Fig01}
\end{figure}

The simulations were performed using the interactive IDL tool \texttt{GS Simulator}. This tool allows one to change the shape and orientation of the flaring loop, choose the parameters of the magnetic field, thermal plasma, and energetic electrons, and calculate the brightness maps of GS emission. In the model used, the magnetic field is produced by a dipole located below the solar surface. The loop is formed by a set of field lines such that at the loop top it has a circular cross section with a given radius; near the footpoints, the loop becomes narrower and non-circular due to conservation of the magnetic flux. As shown in Fig. \ref{Fig01}, which pictures an actual implementation of such geometry in the \texttt{GS Simulator} tool, the user is allowed to freely rotate the dipole loop model in any direction, so that to obtain an arbitrary line of sight relative to the dipole's central plane.

The magnetic model \sout{definitions and} geometry and adjustable parameters are described in detail in Appendix. The source code and documentation of the \texttt{GS Simulator} tool are provided as an online supplement.

\subsection{Flaring loop model and simulation method}

\begin{figure*}
\centerline{%
\includegraphics{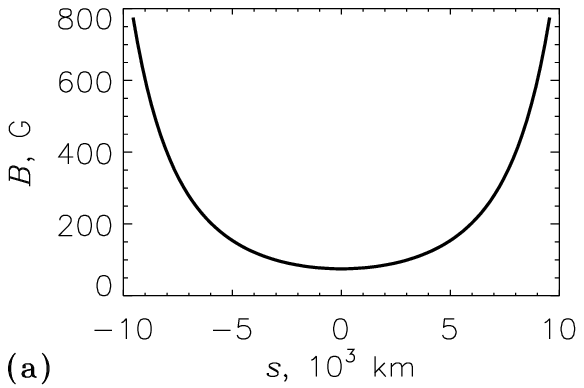}%
\includegraphics{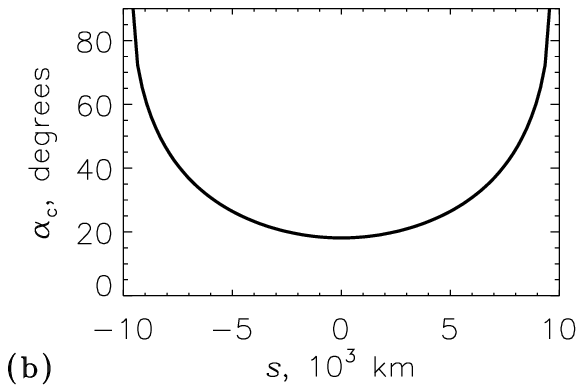}%
\includegraphics{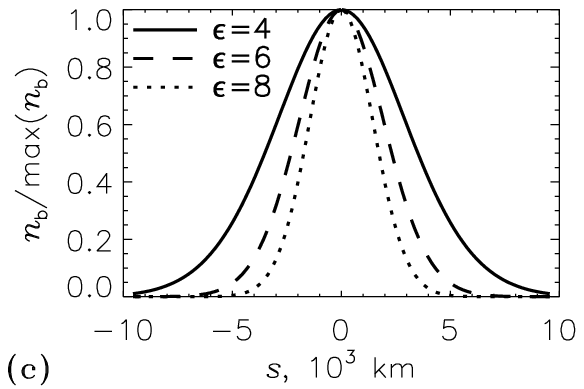}%
}
\caption{Magnetic field strength (a), loss-cone boundary (b), and relative density of the energetic electrons (c) vs. coordinate along the loop. All the values correspond to the loop axis.}
\label{Fig02}
\end{figure*}

To keep the range of options manageable within one paper, we restrict in the present study the flexibility of the model parameters in several ways. We consider only a simple case of  symmetric dipole magnetic loops, i.e. \sout{$\varphi_0=0$} with the dipole perpendicular to a local vertical.  The adopted geometry (visualized via a few reference magnetic field lines demarcating the  surface of the magnetic loop) for two different loop orientations is shown in Fig. \ref{Fig01}. Dependence of the magnetic field strength at the loop axis on the distance from the loop top along the field line is shown in Fig. \ref{Fig02}a.

We assume that the density and temperature of the thermal plasma component within the loop are constant (since the plasma in flaring loops is often heated up to the temperatures of $\gtrsim 10^7$ K, the corresponding barometric scale heights far exceed the loop heights, so the density variations with height can be neglected). Parameters of the energetic electrons can either be constant or vary with the distance from the loop top (see below). By using the above assumptions, we construct a 3D model of the flaring loop (which is observed at a given direction) with all the source parameters depending on the Cartesian coordinates $(x, y, z)$, where $x$ and $y$ are the coordinates in the image plane, and $z$ is the coordinate along the line of sight. The source volume is divided onto a number of volume elements (voxels); each voxel is considered to be quasi-homogeneous. The radio brightness map (or the observed emission intensity as a function of 2D coordinates $x$ and $y$) at a given frequency $f$ is calculated by numerical integration of the radiation transfer equation
\begin{multline}\label{rt}
\frac{\mathrm{d}I_{\mathrm{L, R}}(f, x, y, z)}{\mathrm{d}z}=j_{\mathrm{L, R}}(f, x, y, z)\\
-\varkappa_{\mathrm{L, R}}(f, x, y, z)I_{\mathrm{L, R}}(f, x, y, z)
\end{multline}
along all selected lines of sight. In Eq. (\ref{rt}), $I_{\mathrm{L}}$ and $I_{\mathrm{R}}$ are the spectral intensities of the left- and right-polarized emission components, respectively, $j_{\mathrm{L}}$ and $j_{\mathrm{R}}$ are the corresponding GS emissivities, and $\varkappa_{\mathrm{L}}$ and $\varkappa_{\mathrm{R}}$ are the absorption coefficients. We use the ``weak coupling'' model, i.e., the left- and right-polarized emission components propagate independently. Left-polarized emission corresponds to either ordinary or extraordinary magnetoionic mode, depending on the magnetic field direction; respectively, right-polarized emission corresponds to the opposite mode. The plasma emissivities $j_{\mathrm{O, X}}$ and absorption coefficients $\varkappa_{\mathrm{O, X}}$ for the ordinary and extraordinary modes accounting for both GS and free-free contributions at each voxel are calculated using fast GS codes developed by \citet{Fl_Kuzn_2010}. Outside the flaring loop, the emission propagates like in a vacuum.

The loop orientation is described in general by three Euler angles; however, since rotation of the loop around $z$ axis results simply in the same rotation of the brightness maps, variation of two angles only is considered. We adopt the loop to be  located at the solar equator; in this case, the loop orientation is characterized by the angle $\psi$ between the magnetic dipole and the equatorial plane and by the longitude $\lambda$.

The energetic electrons are described by the distribution function $F$ in a factorized form: $F(E, \mu)=u(E)g(\mu)$, where $E$ is the electron kinetic energy, $\mu=\cos\alpha$, and $\alpha$ is the electron pitch-angle (the angle between the particle velocity and the local magnetic field vectors). The tool allows using a number of different model electron distribution functions, including thermal, power-law, broken power-law, a power-law matched to a maxwellian core (so-called, thermal/nontermal, TNT, distribution), kappa distribution, etc. In particular, use of TNT distribution allows one to consistently take into account both GS (nonthermal) and gyroresonant (GR, thermal) \sout{components} contributions to the emission and absorption. For our modeling, however, it has a disadvantage that two distributions with different anisotropies must be matched, which eventually increases the number of model parameters needed to be varied in the modeling. For this reason, we assume that the electrons have a power-law energy spectrum $u(E)\sim E^{-\delta}$ in the energy range $E_{\min}<E<E_{\max}$ and so neglect the GR contribution entirely. However, we make an assessment of the GR \sout{effect} contribution by considering the TNT distribution in the on-line supplement album. Although the effect of GR absorption is modest for our adopted source model and parameters, it does have \sout{some} a noticeable imprint on the spectrum and polarization \citep[sf,][]{Preka_Alis_1992} from the lower parts of the loop legs, see Figs.~29-30 of the on-line album.

The pitch-angle distribution can be either isotropic or  a loss-cone described by the model function
\begin{equation}\label{lc}
g(\mu)\sim\left\{\begin{array}{ll}
1, & \textrm{for}~|\mu|<\mu_{\mathrm{c}},\\
\displaystyle\exp\left[-\frac{(|\mu|-\mu_{\mathrm{c}})^2}{\Delta\mu^2}\right], & \textrm{for}~|\mu|\ge\mu_{\mathrm{c}},
\end{array}\right.
\end{equation}
where $\mu_{\mathrm{c}}=\cos\alpha_{\mathrm{c}}$, $\alpha_{\mathrm{c}}$ is the loss-cone boundary, and the parameter $\Delta\mu$ determines the sharpness of this boundary. The loss-cone boundary is adopted to exactly follow the transverse adiabatic invariant

\begin{equation}\label{ac}
\sin^2\alpha_{\mathrm{c}}=\frac{B}{B_{\mathrm{f}}},
\end{equation}
where $B$ and $B_{\mathrm{f}}$ are the magnetic fields at a given point and at the loop footpoint, respectively. Dependence of the loss-cone boundary on the coordinate along the loop is shown in Fig. \ref{Fig02}b; this parameter equals $90^{\circ}$ at the footpoint and decreases with height, so that the distribution becomes closer to the isotropic one. Since within the adopted model the anisotropy is only strong at and around the footpoints, while becomes much weaker across most of the magnetic loop, we call this ``a moderate anisotropy''.

We consider both the homogeneous spatial distribution of the energetic electrons (when their number density $n_{\mathrm{e}}$ is constant) and the case when  energetic electrons are accumulated at the loop top \citep{melnikov_etal_2002}. The inhomogeneous distribution of the particles along the loop is described by the following model function:
\begin{equation}\label{ne}
n_{\mathrm{e}}\sim\exp\left[-\epsilon^2(\varphi-\pi/2)^2\right],
\end{equation}
where $\varphi$ is the magnetic latitude (or the angle between the dipole axis and the vector drawn from the dipole center to a given point) and the parameter $\epsilon$ determines the inhomogeneity degree ($\epsilon=0$ corresponds to the homogeneous case). Density profiles of the energetic electrons along the loop for different values of $\epsilon$ are shown in Fig. \ref{Fig02}c.

In the simulations, we use the following parameters of the flaring loop: height $H=10''\simeq 7\,270$~km from the base of the corona, dipole depth below the base of the corona $D=6''\simeq4\,360$~km (so that the distance between the footpoints $\Delta\simeq11.5''\simeq 8\,400$~km), radius at the top $R_{\mathrm{t}}=2''\simeq1\,450$~km, and magnetic field at the top $B_{\mathrm{t}}=75$ G (that results in the footpoint magnetic field of $B_{\mathrm{f}}\simeq 800$ G). Figures \ref{Fig01}-\ref{Fig02} correspond to these parameters. Two loop orientations are considered in the paper in some detail: $\psi=60^{\circ}$, $\lambda=20^{\circ}$ (a loop near the center of the solar disk, Fig. \ref{Fig01}a) and $\psi=60^{\circ}$, $\lambda=80^{\circ}$ (a loop at the limb, Fig. \ref{Fig01}b); many more examples are given in the on-line album. The thermal plasma density and temperature are $n_0=10^{10}$ $\textrm{cm}^{-3}$ and $T_0=2\times 10^7$ K, respectively. The energetic electrons have the power-law index $\delta=4$, cutoff energies $E_{\min}=100$ keV and $E_{\max}=10$ MeV, and the loss-cone boundary width $\Delta\mu=0.2$. Thus, in each loop orientation, the variable parameters are: type of the pitch-angle distribution (isotropic or loss-cone), the number density of the energetic electrons $n_{\mathrm{e}}$, and the inhomogeneity parameter $\epsilon$.

\begin{figure*}
\centerline{\includegraphics{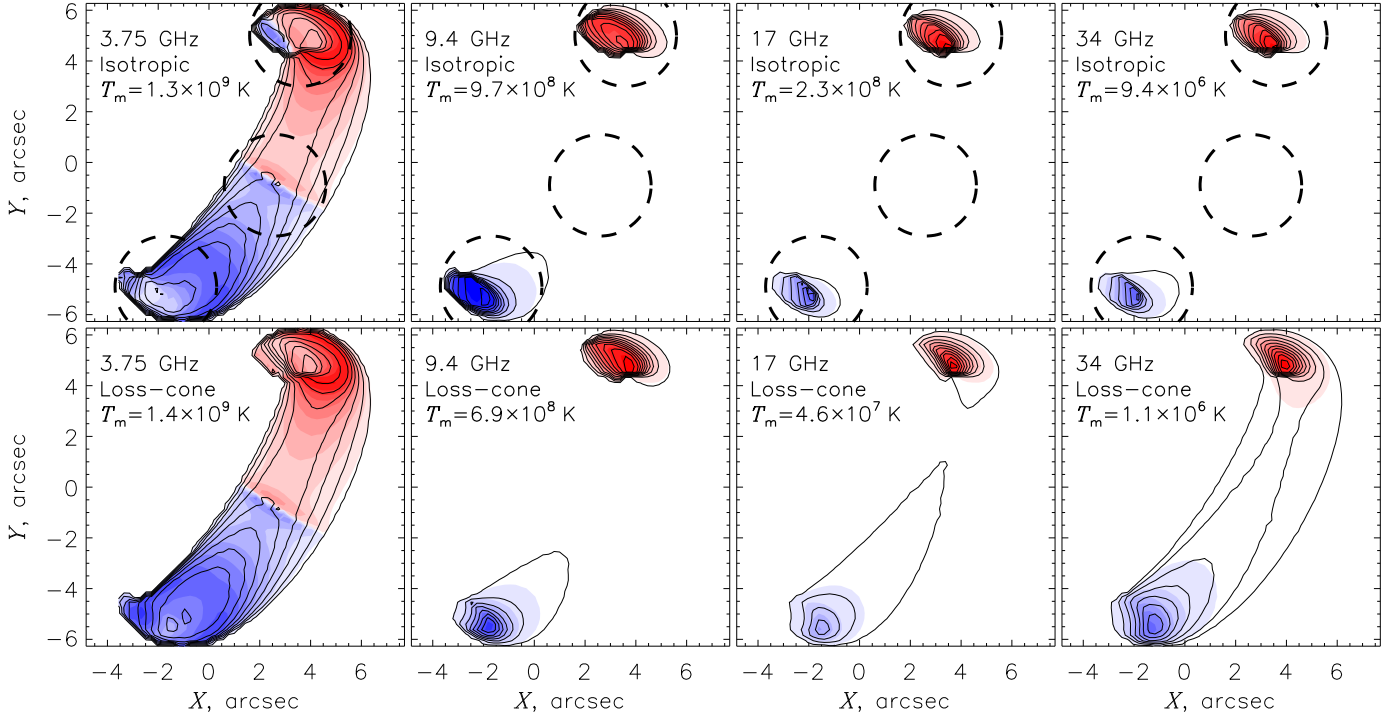}}
\caption{Radio brightness maps for a loop located near the center of the solar disk, for the isotropic distribution (top row) and loss-cone distribution (bottom row); the north is up and the west is to the right. Concentration of the accelerated electrons is assumed to be constant along the loop. Solid lines are the intensity contours which are evenly distributed between zero and the maximum brightness temperature $T_{\mathrm{m}}$ (the corresponding temperatures are given in each panel). Color shades (see online version of the journal) represent the circular polarization (Stokes $V$ normalized by the absolute value of $V$ peak); red and blue correspond to the right and left circular polarizations, respectively.}
\label{Fig03}
\end{figure*}
\begin{figure*}
\centerline{\includegraphics{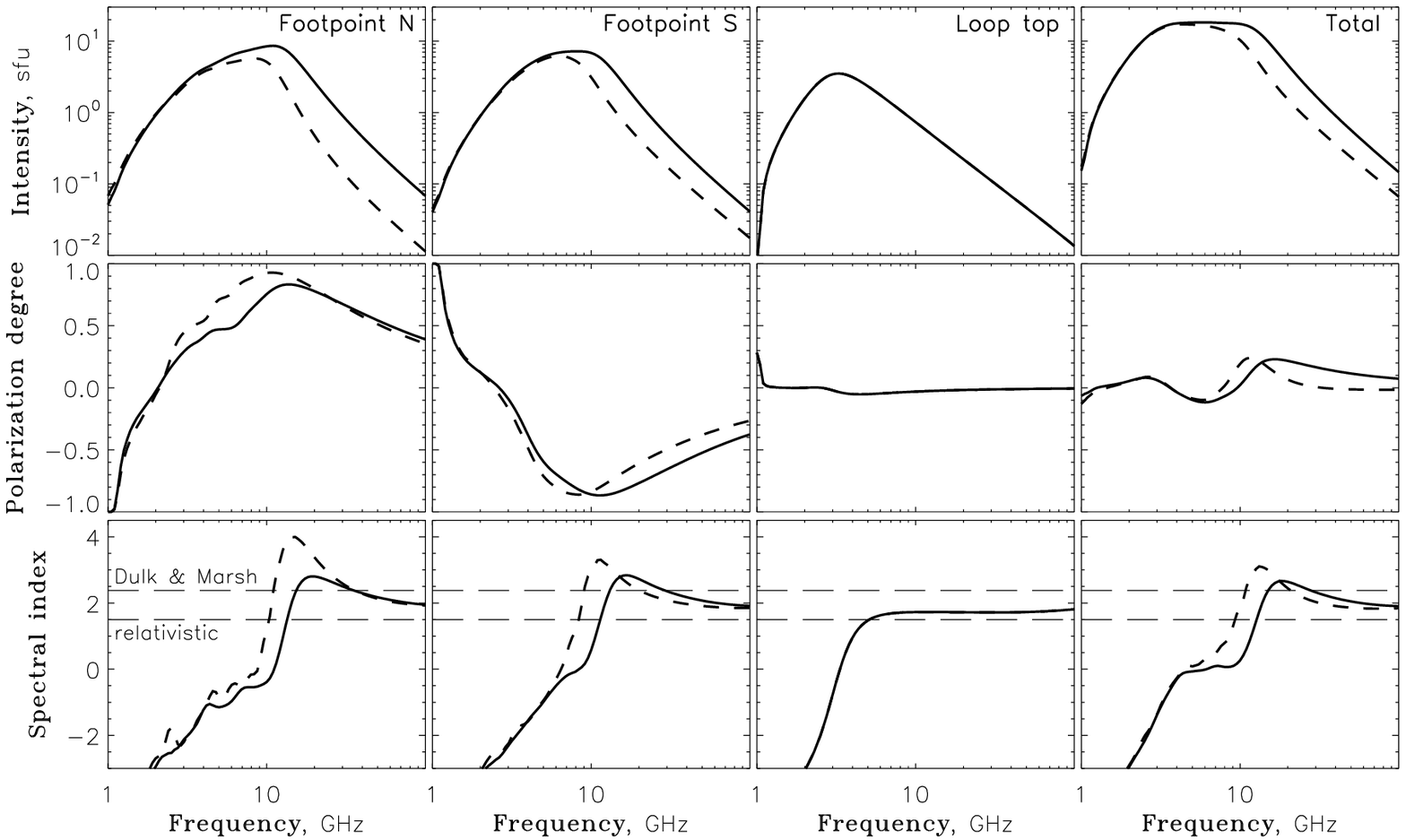}}
\caption{Emission intensity, degree of polarization, and spectral index vs. frequency for the loop shown in Fig. \protect\ref{Fig03}. The columns (from left to right) correspond to: northern footpoint source, southern footpoint source, loop-top source, and the emission from the entire loop (spatially unresolved). Solid lines: isotropic distribution; dashed lines: loss-cone distribution. The regions taken to calculate the spatially resolved spectra are indicated in Fig. \protect\ref{Fig03} by thick dashed circles.}
\label{Fig04}
\end{figure*}

\section{Simulation results}
\subsection{Effect of the electron anisotropy}\label{aniso}
Firstly, we consider effect of  electron anisotropy on the GS emission. Figure \ref{Fig03} shows the brightness maps at four frequencies for the loop located near the center of the solar disk. The displayed Stokes parameters are: $I=I_{\mathrm{R}}+I_{\mathrm{L}}$ and $V=I_{\mathrm{R}}-I_{\mathrm{L}}$. The isotropic (top row) and loss-cone (bottom row) pitch-angle distributions are considered. In both cases, the number density of the energetic electrons $n_{\mathrm{e}}$ is constant and equals $3\times 10^6$ $\textrm{cm}^{-3}$. Figure \ref{Fig04} shows the spatially resolved emission spectra for the footpoint and loop top sources (obtained by summation over the circled pixels) as well as the total emission (obtained by summation over all pixels). The figure also shows the degree of polarization $\eta$ and spectral index $\delta_{\mathrm{r}}$ which are defined as
\begin{equation}\label{es}
\eta=\frac{V}{I},\qquad
\delta_{\mathrm{r}}=-\frac{f}{I}\frac{\mathrm{d}I}{\mathrm{d}f}.
\end{equation}

From the spatially resolved spectra, one can see that the influence of the electron anisotropy is most important at the footpoints. In the optically thin frequency range, the emission from the loss-cone distribution is lower than that from the isotropic electrons by a factor of about 2-6. This reflects the fact that the GS radiation is emitted mainly in the direction of the electron velocity. Near the footpoints, the electrons with the loss-cone distribution are concentrated around the pitch-angle of $90^{\circ}$ whereas the magnetic field (in the adopted geometry) is nearly parallel to the line of sight. This is why the electrons with the loss-cone distribution produce only a relatively weak radiation flux towards the observer. In the optically thick frequency range, the emission intensities from the isotropic and anisotropic distributions are almost the same, although the differences in the degree of polarization and spectral index can be visible. Near the loop top, the loss-cone boundary $\alpha_{\mathrm{c}}$ falls to about $20^{\circ}$ and thus the loss-cone distribution does not noticeably differ from the isotropic one; as a result, these distributions produce almost identical emission.

A complementary way of thinking of the emission is via the radio images at various frequencies (Fig. \ref{Fig03}). At the low frequencies (3.75 GHz), the whole loop is seen on the map (although the footpoints are brighter than the top); the images for the isotropic and anisotropic distributions are very similar. At the higher frequencies, the emission is strongly concentrated at the footpoints. However, since the footpoint emission from the electrons with the loss-cone distribution is weaker than that from the isotropic distribution, the difference between the footpoints and the loop top is smaller as well. Therefore the emission from the anisotropic electrons is more evenly distributed along the loop.

\begin{figure*}
\centerline{\includegraphics{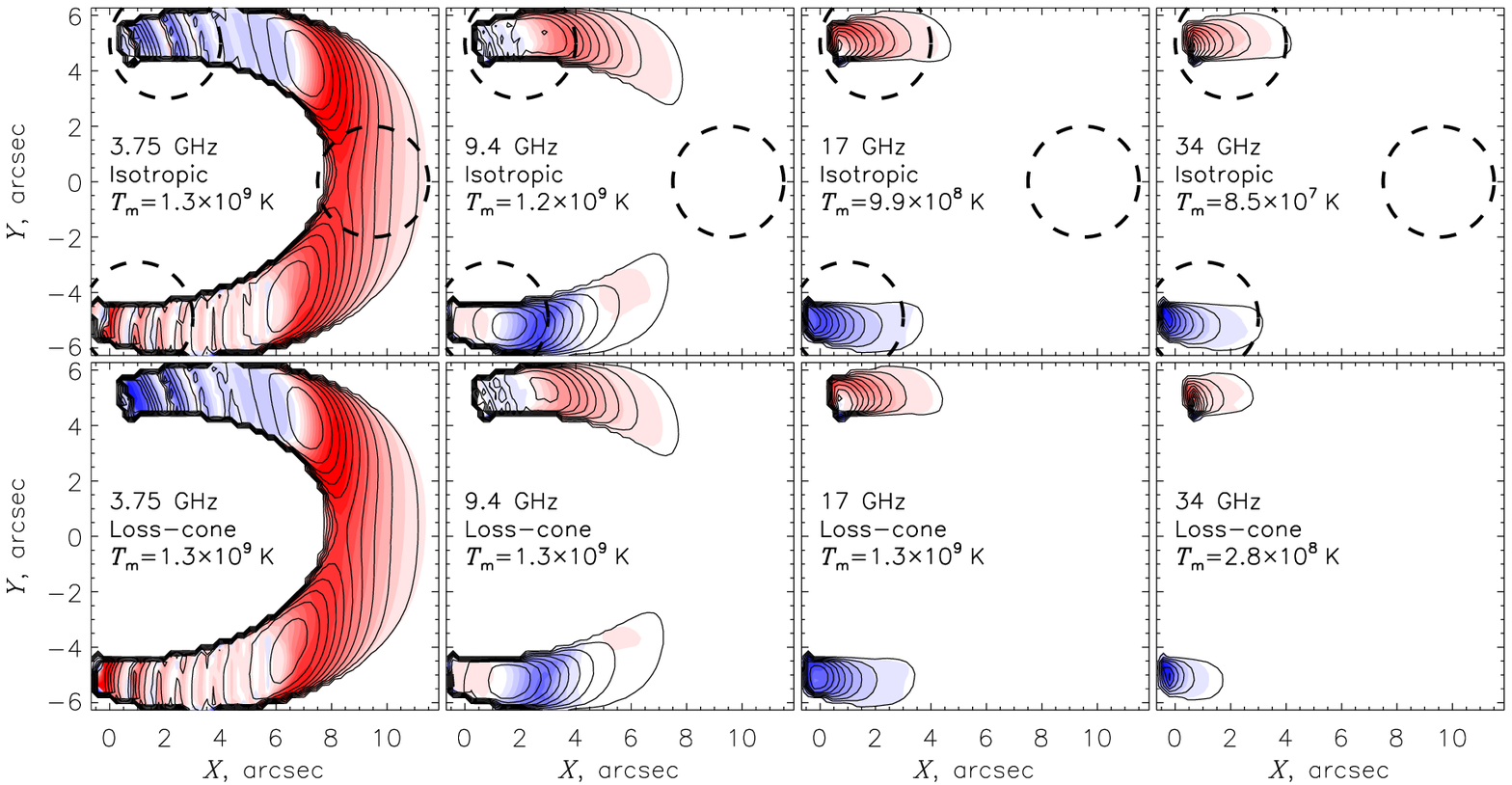}}
\caption{The same as in Fig. \protect\ref{Fig03}, for a loop located near the solar limb.}
\label{Fig05}
\end{figure*}
\begin{figure*}
\centerline{\includegraphics{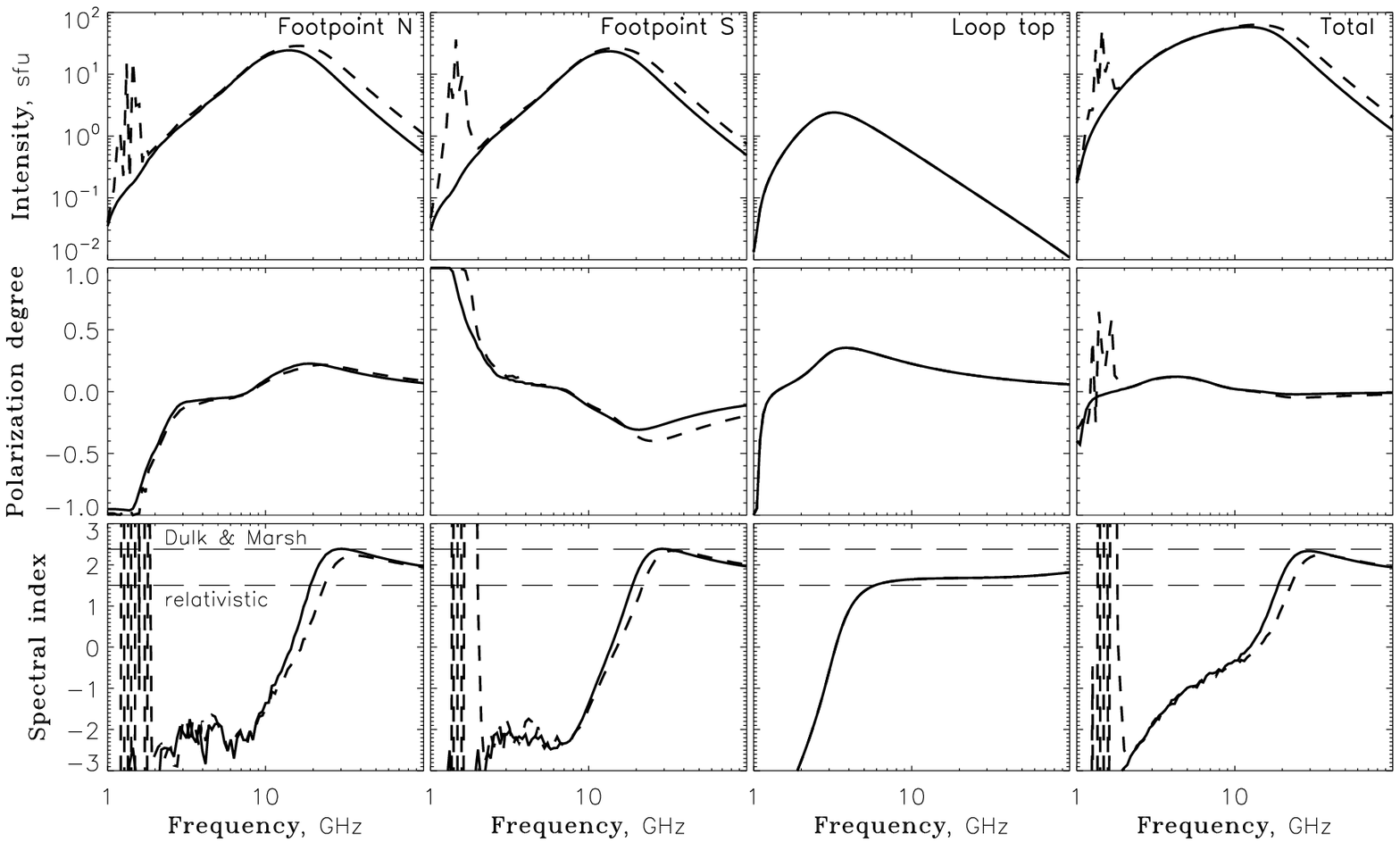}}
\caption{The same as in Fig. \protect\ref{Fig04}, for the loop shown in Fig. \protect\ref{Fig05} (located near the solar limb). The regions taken to calculate the spatially resolved spectra are indicated in Fig. \protect\ref{Fig05} by thick dashed circles.}
\label{Fig06}
\end{figure*}

Figures \ref{Fig05}-\ref{Fig06} show the brightness maps and emission spectra for the loop located near the limb (all other parameters are the same as in Figures~\ref{Fig03}-\ref{Fig04}). One can see that now the electrons with the loss-cone distribution produce stronger footpoint emission in the optically thin frequency range than the isotropic distribution does (due to the same reasons as discussed above). In the optically thick frequency range, the considered characteristics of the emission are very similar for both isotropic and anisotropic cases except the effects of the individual cyclotron harmonics that are much more pronounced for the anisotropic distribution, which is discussed in Section \ref{harm} in greater detail. The loop top emissions from the isotropic and loss-cone electron distributions are almost the same.

In the brightness maps (Fig. \ref{Fig05}), the maximum of the emission at 3.75 GHz is located between the loop top and the footpoints; the images for the isotropic and anisotropic distributions are very similar. At the higher frequencies, the emission is concentrated at the footpoints. In contrast to the previous case (Fig. \ref{Fig03}), the footpoint sources are now more compact for the loss-cone electron distribution; this effect becomes more pronounced with the frequency increase.

Changing the pitch-angle distribution from the isotropic one to the loss cone results in a shift of the spectral peak of the footpoint emission towards lower frequencies for the loop located near the disk center, and towards higher frequencies for the loop near the limb (which could be observationally addressed via center-to-limb variation analysis of the spatially resolved footpoint spectra). This spectral peak variation affects the emission spectral index around the peak. With an increasing frequency (in the optically thin range), the spectral indices of the emission from the isotropic and anisotropic electrons approach each other and gradually become the same. We have confirmed that at $f\to\infty$, the emission spectral indices asymptotically approach the ultrarelativistic limit $\delta_{\mathrm{rel}}=(\delta-1)/2$ \citep[provided that the high-energy cutoff $E_{\max}\to\infty$,][and neglecting the free-free contribution]{gin65} for both the isotropic and anisotropic electron distributions. However, in the frequency range of $\sim 10-100$ GHz, the emission spectral index is strongly affected by the magnetic field inhomogeneity in the source; as a result, the spectral index varies with frequency and can be noticeably different at the different footpoints and the loop top. At the frequencies $\gtrsim 100$ GHz, the spectrum is affected  by the high-energy cutoff of the electron distribution (note that Figs. \ref{Fig03}-\ref{Fig06} are calculated for $E_{\max}=10$ MeV) which makes the spectrum steeper (the spectral index increases with frequency); this effect is clearly visible for the loop-top emission where the magnetic field inhomogeneity is the lowest. On the other hand, at these and higher frequencies the free-free contribution can make the spectrum flatter (see below examples of such flattening).

The  loss-cone anisotropy of the fast  electrons has opposite effect on the GS emission for the loops located near the solar disk center and the limb. Thus, for some intermediate longitude, the anisotropy effect should be minimal. We have found that this occurs at the longitude of $\lambda\simeq 60^{\circ}$.

\begin{figure*}
\centerline{\includegraphics{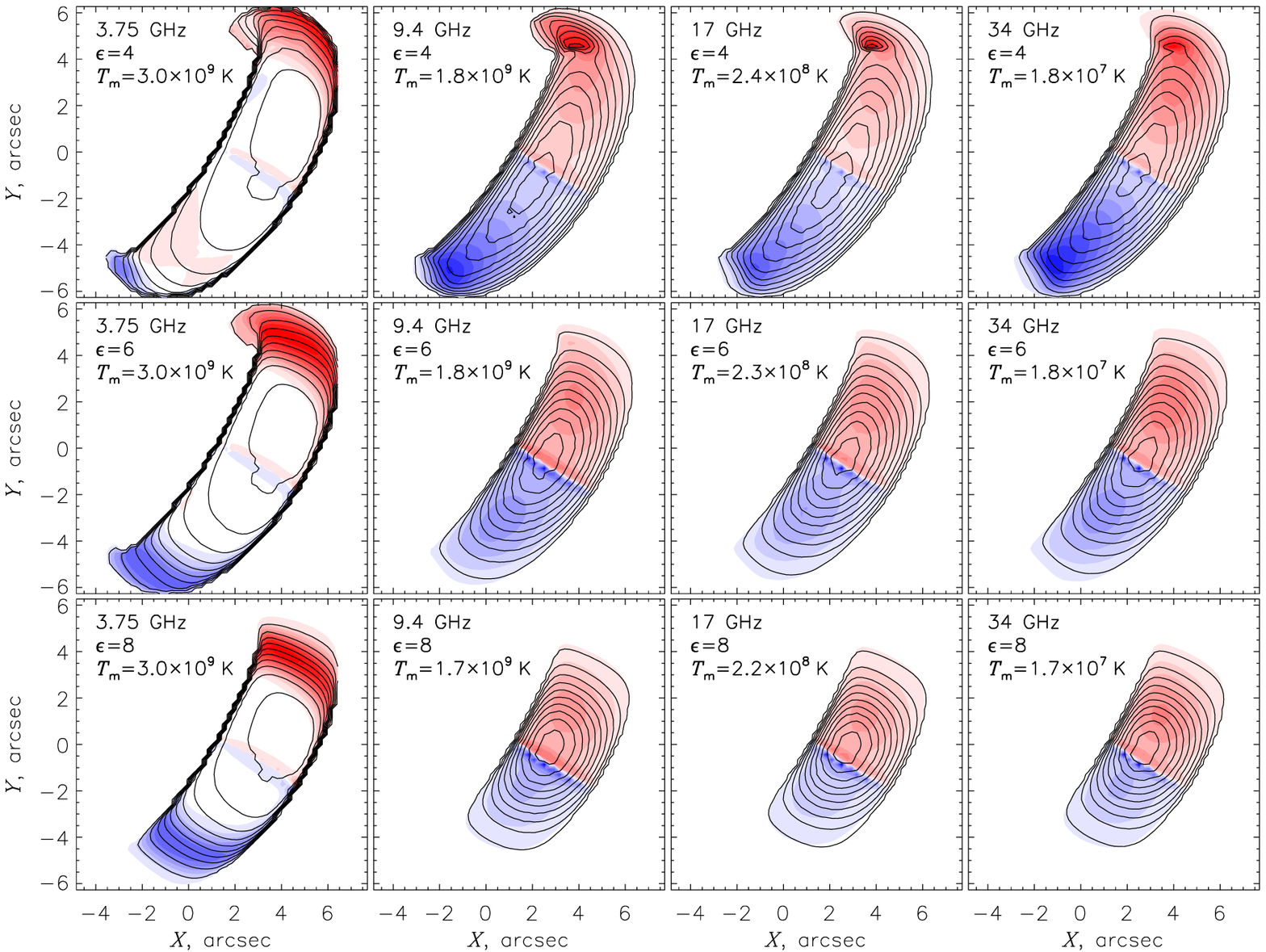}}
\caption{Radio brightness maps for a loop located near the center of the solar disk for the different distributions of the accelerated electrons along the loop. The accelerated electrons are assumed to have the loss-cone distribution.}
\label{Fig07}
\end{figure*}
\begin{figure*}
\centerline{\includegraphics{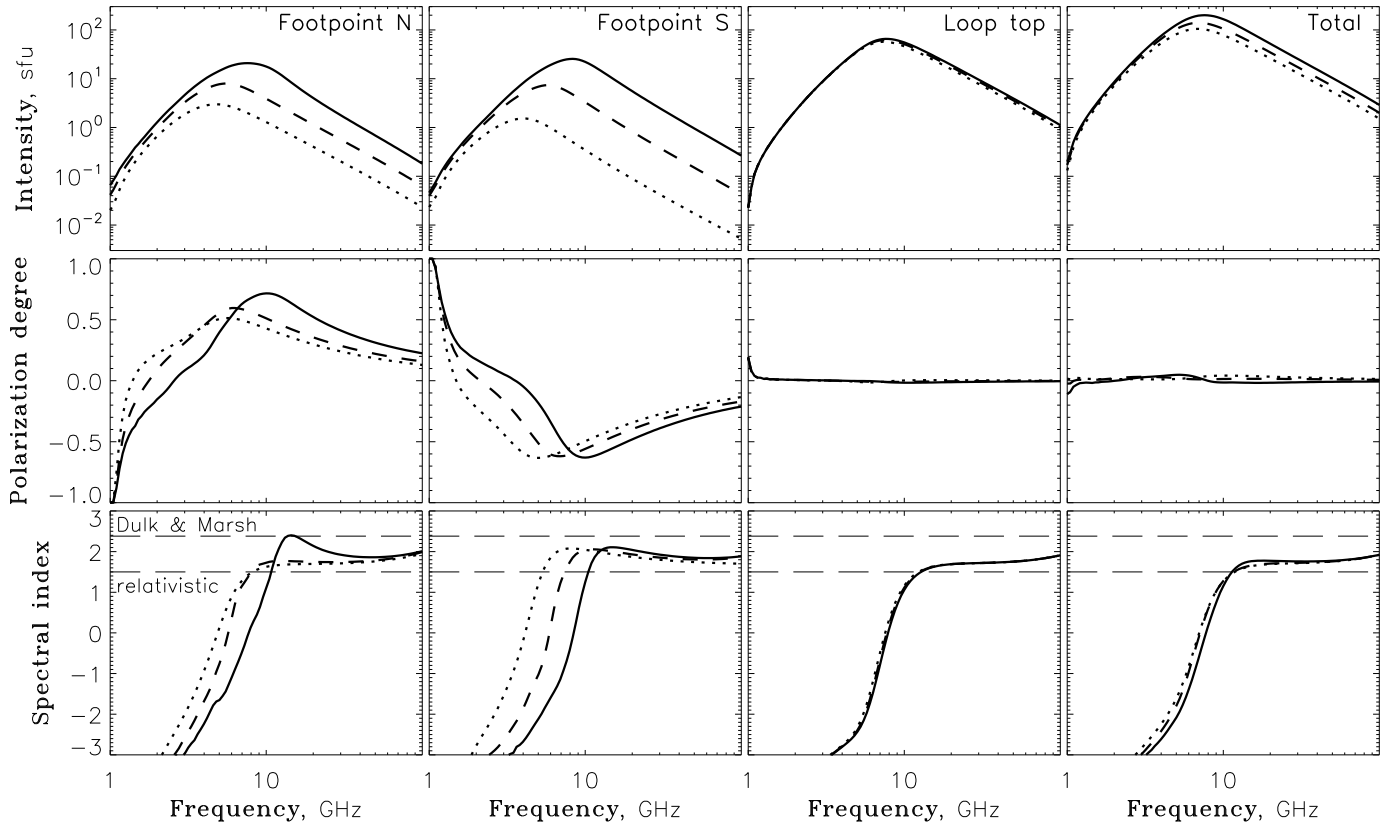}}
\caption{Emission intensity, degree of polarization, and spectral index vs. frequency for the loop models shown in Fig. \protect\ref{Fig07}. Solid lines: $\epsilon=4$; dashed lines: $\epsilon=6$; dotted lines: $\epsilon=8$. The regions taken to calculate the spatially resolved spectra are indicated in Fig. \protect\ref{Fig03} by thick dashed circles.}
\label{Fig08}
\end{figure*}
\begin{figure*}
\centerline{\includegraphics{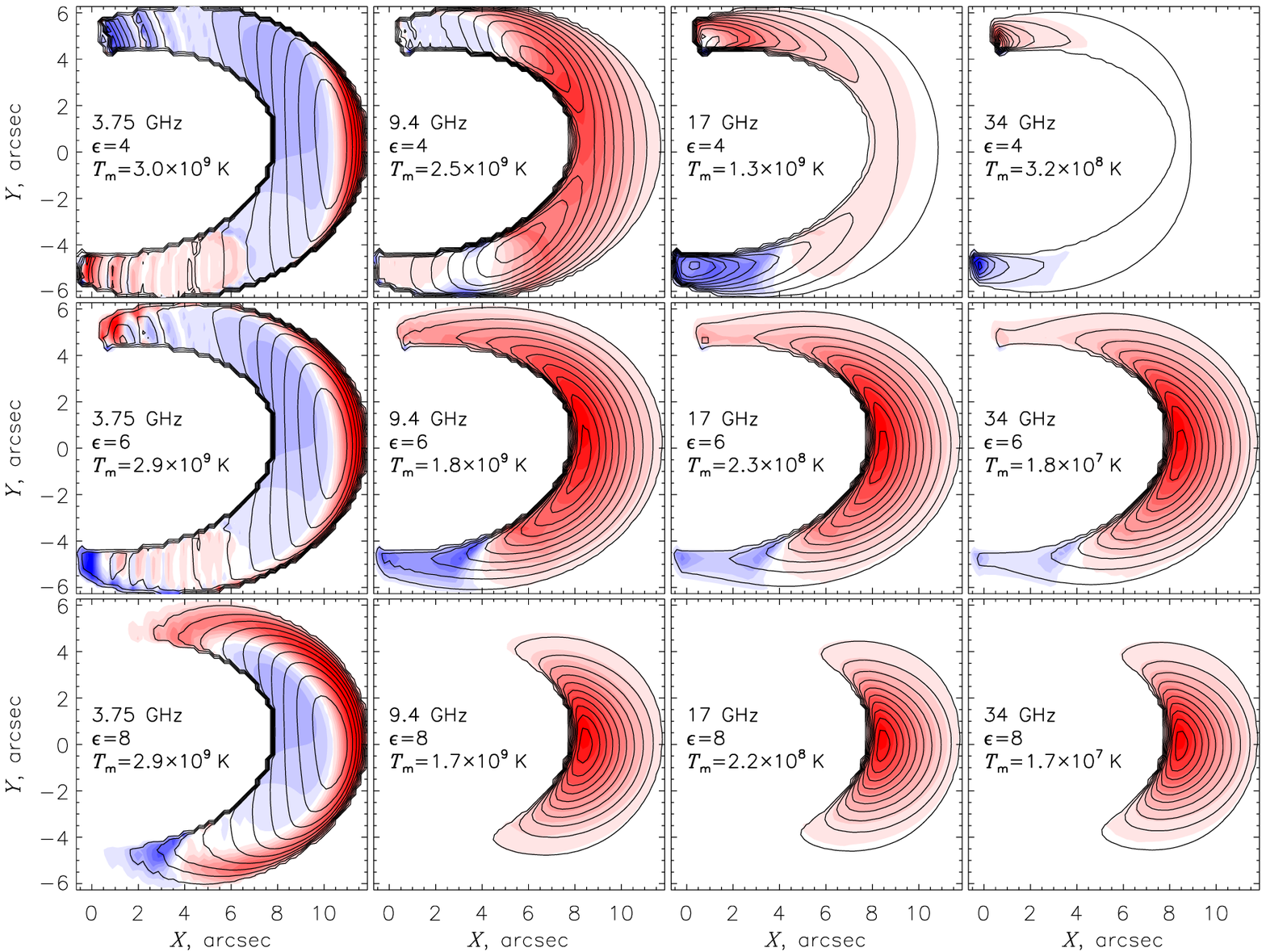}}
\caption{The same as in Fig. \protect\ref{Fig07}, for a loop located near the solar limb.}
\label{Fig09}
\end{figure*}
\begin{figure*}
\centerline{\includegraphics{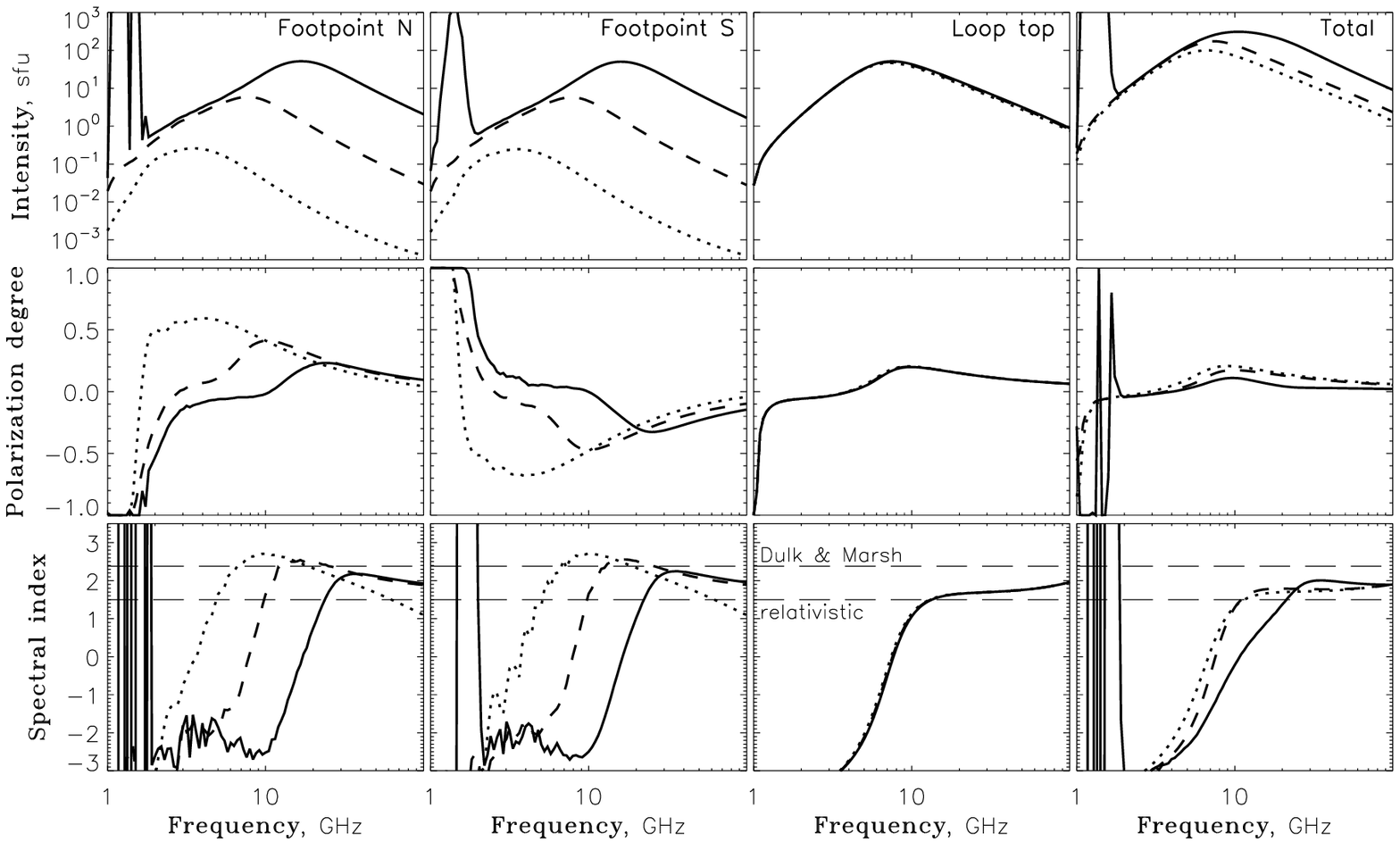}}
\caption{The same as in Fig. \protect\ref{Fig08}, for the loop models shown in Fig. \protect\ref{Fig09} (the loop is located near the solar limb). The regions taken to calculate the spatially resolved spectra are indicated in Fig. \protect\ref{Fig05} by thick dashed circles.}
\label{Fig10}
\end{figure*}

\subsection{Effect of the inhomogeneous electron distribution}\label{inhom}
Figures \ref{Fig07}-\ref{Fig10} show the brightness maps and spatially resolved emission spectra for the models with the inhomogeneous distribution of the energetic electrons along the flaring loop (the particles are accumulated at the loop top). The number density of the energetic electrons at the loop top is taken to be $n_{\mathrm{t}}=2.8\times 10^8$ $\textrm{cm}^{-3}$. The density profiles of the energetic electrons are described by Equation (\ref{ne}) with $\epsilon=4$, 6, and 8, which corresponds to the ratios of the footpoint and loop-top number densities of $n_{\mathrm{f}}/n_{\mathrm{t}}=1.1\times 10^{-2}$, $3.6\times 10^{-5}$, and $1.3\times 10^{-8}$, respectively (for $\epsilon=4$, we obtain the footpoint number density of $n_{\mathrm{f}}=3\times 10^6$ $\textrm{cm}^{-3}$, like in the previous Section). In all cases, the loss-cone distribution of the energetic electrons over the pitch-angle is used.

Figures \ref{Fig07}-\ref{Fig08} correspond to the loop located near the center of the solar disk. Since the number density of the energetic electrons at the loop top is assumed to be constant, the loop-top emission is almost independent on the parameter $\epsilon$. In the footpoints, the increase of $\epsilon$ results in a decrease of the number density of the energetic electrons and, consequently, in a decrease of the emission intensity in the optically thin frequency range; the spectral peak shifts towards lower frequencies. In the brightness maps, the intensity maximum is always located near the loop top. For $\epsilon=4$, at the frequency of 3.75 GHz, the whole loop is seen with relatively weak brightness variations along it. At higher frequencies and/or with increasing $\epsilon$, the visible emission source becomes more compact.

Figures \ref{Fig09}-\ref{Fig10} show the brightness maps and emission spectra for the loop located near the limb. Like in the previous case, the loop-top emission is independent on the parameter $\epsilon$ while the footpoint emission intensity (in the optically thin frequency range) decreases with increasing $\epsilon$. Distributions of the radio brightness along the loop can be qualitatively different for the different frequencies and inhomogeneity models: for $\epsilon=4$, the emission maximum at 3.75 GHz is located at the loop top and gradually shifts towards the footpoints with increasing frequency. For $\epsilon=6$ and 8, the intensity maximum is always located at the loop top, but the visible emission source becomes more compact with increasing frequency. Thus at high frequencies ($\gtrsim 15$ GHz), a brightness map will be dominated by either two footpoint sources (for the loops with a moderate accumulation of the particles at the top, up to $n_{\mathrm{t}}/n_{\mathrm{f}}\simeq 100$) or one loop-top source (for the loops with higher inhomogeneity). We note an interesting behavior of the polarization map: for the loop-like optically thin sources (at 17 or 34~GHz) in Figure~\ref{Fig09} most of the image area is dominated by the RCP sense, which is indicative of the same direction of the  line-of-sight magnetic field component throughout most of the loop area. Similar unipolar polarization patterns are highly typical for the NoRH data on the limb flares.     If the concentration of the electrons at the loop top develops at the course of flare due to electron transport \citep{Lee_Gary_Zirin_1994, melnikov_etal_2002, Fl_2005n}, this process of the electron inhomogeneity build-up can be roughly described in terms of the $\epsilon$ increase, which implies the apparent brightness peak motion (at the optically thin frequencies) from footpoints to the looptop in agreement with observations of some limb flares  \citep{Tzatzakis_etal_2008, Reznikova_etal_2009}.

In all the models, the emission spectral index at the loop top behaves like that for a homogeneous source: the index is nearly constant ($\delta_{\mathrm{r}}\simeq 1.7-1.8$) in the frequency range of 20-50 GHz (for both loop orientations), and increases at higher frequencies due to the high-energy cutoff of the electron spectrum. At the footpoints, an increase of the parameter $\epsilon$ results in a decrease of the spectral peak frequency. With an increasing frequency (in the optically thin range), the spectral indices for all values of $\epsilon$ become the same (if the free-free emission is negligible, see below). However, in contrast to the loop top, the spectral index of the footpoint emission is also affected by the source inhomogeneity. For the loop located near the disk center, we can notice an asymmetry of the emission spectra produced at the different footpoints, which is caused by a difference of the viewing angles. For the loop located near the limb, the high-frequency spectral index of the footpoint emission for the model with $\epsilon=8$ differs noticeably from the indices for $\epsilon=4$ and 6. This is caused by the contribution of the free-free emission of the thermal electrons which becomes dominant under these conditions: at large frequencies 
the spectral index of the free-free emission goes to zero.

Changing the pitch-angle distribution of the energetic electrons affects the spatially resolved emission spectra in the same way as discussed in the previous Section: in comparison to the iso\-tro\-pic distribution, the loss-cone one provides stronger footpoint emission for a loop near the limb and weaker footpoint emission for the loop near the disk center, while the loop-top emission remains almost unchanged. Since in the considered inhomogeneous models the loop-top emission source often dominates the images, the effect of the electron ani\-so\-tro\-py on the brightness maps is only moderate.

\begin{deluxetable}{lccccc}
\tablecolumns{6}
\tablewidth{0pt}
\tablecaption{Spectral parameters of the total (spatially unresolved) emission.\label{Tab1}}
\tablehead{\colhead{$g(\mu)$} & \colhead{$\epsilon$} & \colhead{$f_{\mathrm{peak}}$, GHz} & \colhead{$\delta_{15}$} & \colhead{$\delta_{17-34}$} & \colhead{$\delta_{50}$}}
\startdata
\sidehead{\it Loop location: near the disk center}
Isotropic & 0 & 6.03 & 2.40 & 2.46 & 2.05 \\
Loss cone & 0 & 4.37 & 2.98 & 2.21 & 1.86 \\
Isotropic & 4 & 7.59 & 1.75 & 1.82 & 1.79 \\
Loss cone & 4 & 7.59 & 1.76 & 1.77 & 1.77 \\
Loss cone & 6 & 6.92 & 1.66 & 1.72 & 1.76 \\
Loss cone & 8 & 6.92 & 1.64 & 1.71 & 1.75 \\
\sidehead{\it Loop location: near the limb}
Isotropic & 0 & 12.02 & 0.57 & 2.10 & 2.14 \\
Loss cone & 0 & 13.80 & 0.13 & 1.67 & 2.16 \\
Isotropic & 4 & 10.47 & 0.94 & 1.87 & 1.92 \\
Loss cone & 4 & 10.47 & 0.79 & 1.72 & 1.95 \\
Loss cone & 6 & 7.24 & 1.77 & 1.78 & 1.78 \\
Loss cone & 8 & 6.61 & 1.66 & 1.71 & 1.76 \\
\enddata
\tablecomments{The values $\delta_{15}$ and $\delta_{50}$ are the exact spectral indices at 15 and 50 GHz, respectively; $\delta_{17-34}$ is an approximate spectral index calculated using two points at 17 and 34 GHz; for comparison $\delta_{\mathrm{rel}}=1.5$ while $\delta_{\mathrm{D-M}}=2.38$. For the models with $\epsilon=0$ and $\epsilon\neq 0$, the number densities of the energetic electrons at the loop top are different (see text).}
\end{deluxetable}

\subsection{Spatially integrated spectra}
In this Section, we consider the total (spatially integrated) emission from the flaring loop. Although in the view of the currently operational and soon to be  available radio instruments, considering the total radiation may look old-fashioned, we feel that this still makes sense for the following reasons. Most of the historically accumulated databases and corresponding statistical studies are done based on the total power observations \citep[e.g.,][]{Guidice_Castelli_1975, Nita_etal_2004}. The total power data are simpler manageable as they can be  easily visualized by dynamic spectra and characterized by only a few simple numbers, such as spectral indices, rise and decay times, peak flux and frequency \citep{Nita_etal_2004}. In particular, the corresponding high-frequency spectral index is widely used to evaluate the fast electron energy spectral index \citep{dul82}. From this perspective, a statement made by \cite{Simoes_Costa_2010} that such a spectral index is a good measure of the electron spectral index even for anisotropic electron distributions, if confirmed, could be of a great practical value.

The  parameters characterizing the total emission are shown in the right columns in Figs. \ref{Fig04}, \ref{Fig06}, \ref{Fig08}, and \ref{Fig10}. Considering the right top panel in Figure~\ref{Fig04} as a vivid example, one can easily isolate three distinct regions of the spectra---low-frequency part (region I), middle-frequency part (region II), and high-frequency part (region III).  Visual comparison of these total power spectra with the spatially resolved spectra from the footpoints and looptop in Figure~\ref{Fig04} suggests that the low-frequency part is formed primarily at the looptop region with low magnetic field, the high-frequency part in the footpoints where the magnetic field is large, while the middle-frequency part by the entire loop and so related to the magnetic field non-uniformity.

The low-frequency part is known to be formed by the effect of the GS optical thickness and/or the Razin-effect \citep[suppression of the GS emission in a dense background plasma, e.g.,][]{Bastian_etal_1998} possibly accompanied by the free-free absorption in the dense plasma \citep{Bastian_etal_2007}; the slope of the spectrum  can here be quantified by the index of $-2$ or less, see the right bottom panel in Figure~\ref{Fig04}. The high-frequency part is mainly determined by the distribution of fast electrons including the energy spectrum \citep{Bastian_etal_1998} and pitch-angle anisotropy \citep{FlMel_2003b}; note the anisotropy-related difference between the solid (isotropic) and dashed (loss-cone) curves in this panel.

The middle-frequency part is clearly seen in Fig. \ref{Fig04} as it is almost flat (the spectral index is around zero). Solar microwave bursts with flat spectra have been observed for decades. For example, \citet{Ramaty_Petrosian_1972} proposed that the free-free absorption of GS emission can form such spectra, while \citet{Lee_Gary_Zirin_1994} recognized that the electrons trapped in a large dipole magnetic loop can produce the flat radio spectra due to the source non-uniformity in certain parameter regimes. Our modeling confirms the finding made by \citet{Lee_Gary_Zirin_1994}. In addition, we have found that this middle-frequency part is not always flat but can have either negative or positive spectral index, see, e.g., Fig.~\ref{Fig06}, which can be misinterpreted as either region I or III in observations with a limited spectral coverage. In fact, the observations \citep[see, e.g., Fig. 11 in][]{Nita_etal_2004} reveal that the histograms of both low- and high-frequency spectral indices extend to zero implying that both low- and high-frequency spectra can be much flatter than those determined by optical thickness effect or electron energy index, respectively. For practical application, this means that having the spectrum falling with the frequency does not guarantee that its slope is formed by either energetic or angular properties of the electron distribution function but can instead be related to the source non-uniformity.

One of the parameters controlling the shape of the middle-frequency part is the viewing angle: Figure~\ref{Fig06} presents the case when this part grows with frequency, which overall broadens the spectrum peak. Not surprisingly, non-uniform spatial distribution also significantly affect this part of the spectrum. In fact, with an inhomogeneous non-thermal electron spatial distribution (with their concentration at the looptop), the radio spectrum begins to resemble emission from a roughly uniform source, see Figures~\ref{Fig08} and \ref{Fig10}. The reason for this to happen for an \textit{inhomogeneous} source is very simple: with the adopted inhomogeneous electron distribution most of them reside at the looptop, where the spatial variation of the magnetic field and the viewing angle are minor, so we have a situation similar to a uniform source.

Finally, let us consider to what extent we could rely on the high-frequency spectral index in evaluating the fast electron energy spectrum or pitch-angle anisotropy. The obtained spectral behavior of the local spectral index can easily be tracked in the figures, so we do not describe it here in any detail. Instead,  in order to quantify it with measures relevant to available observations, we form a few distinct spectral indices, namely, an index $\delta_{15}$ at 15 GHz relevant to the OVSA spectral range \citep{Nita_etal_2004}, the ``Nobeyama'' index $\delta_{17-34}$ calculated with two well separated frequencies, 17 and 34 GHz \citep[e.g.,][]{Reznikova_etal_2009}, and higher-frequency spectral index $\delta_{50}$ taken at 50 GHz relevant to a number of earlier separated-frequency observations \citep[e.g.,][]{Melnikov_Magun_1998}; the results are gathered in Table~\ref{Tab1}.

One can see, that for the same energy electron distribution, $\delta = 4$, the measured spectral indices are highly different depending on the pitch-angle anisotropy, the viewing angle, spatial inhomogeneity of the fast electron distribution, and the instrument used to measure the spectral index. In fact, the numbers given in Table~\ref{Tab1} are bounded between 0.13 and 2.98, although as has already been said, the ultrarelativistic approximation predicts the emission spectral index of $\delta_{\mathrm{rel}}=(\delta-1)/2$, or $\delta_{\mathrm{rel}}=1.5$ for the adopted value of $\delta=4$, while a widely used approximation proposed by \citet{dul82} predicts the value of $\delta_{\mathrm{D-M}}\simeq 0.90\delta-1.22$, or $\delta_{\mathrm{D-M}}\simeq 2.38$ for $\delta=4$; none of this approximate values is favored by Table~\ref{Tab1}. Stated another way, the radio indices from 0.13 to 2.98 would imply the electron energy index range from 1.26 to 6.96 if the relativistic asymptote is used and from 1.50 to 4.67 if the Dulk-Marsh approximation is applied. Again, none of the values is truly informative in terms of recovering the original electron energy index as there is no unique way of linking the measured radio spectral index with the original electron energy spectral index. This conclusion holds (with narrower scatter of the recovered values, $4.34-5.92$ and $3.21-4.09$, respectively) if we consider only higher-frequency indices $\delta_{17-34}$ and $\delta_{50}$. On the other hand, the OVSA index $\delta_{15}$ is more sensitive to the particle anisotropy, which, thus, can be studied through a forward fit of the radio spectra like in \citet{Fl_etal_2009}.

\begin{figure*}
\centerline{\includegraphics{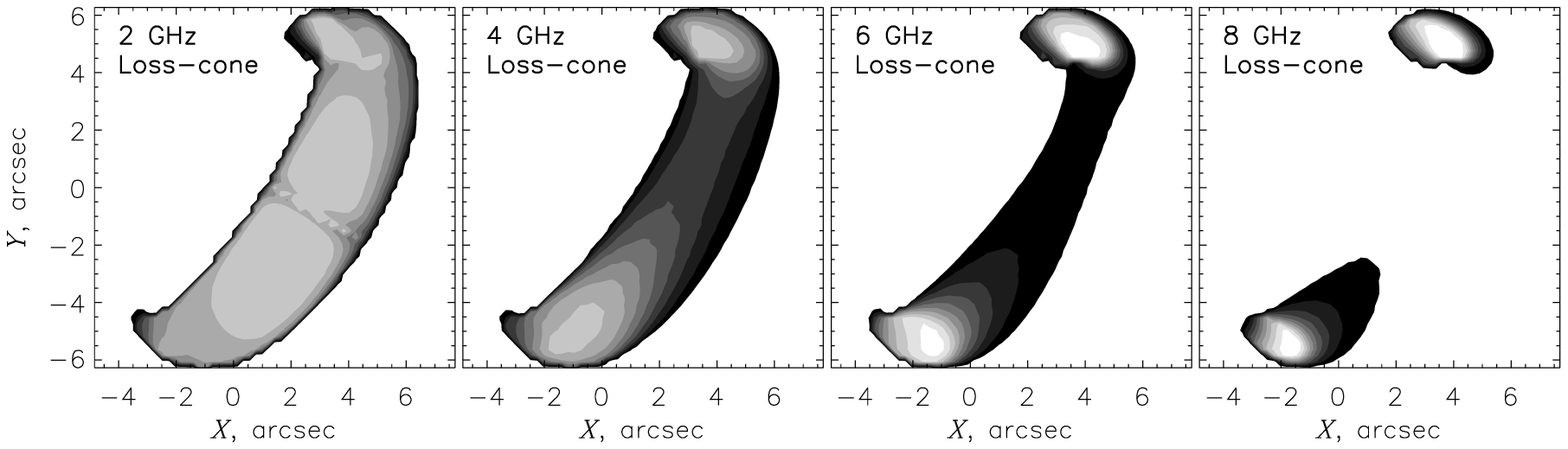}}
\caption{Radio brightness maps (emission intensity) for a loop located near the center of the solar disk. Brighter areas correspond to higher intensity. Number density of the accelerated electrons is constant and their pitch-angle distribution is of the loss-cone type.}
\label{Fig11}
\end{figure*}
\begin{figure*}
\centerline{\includegraphics{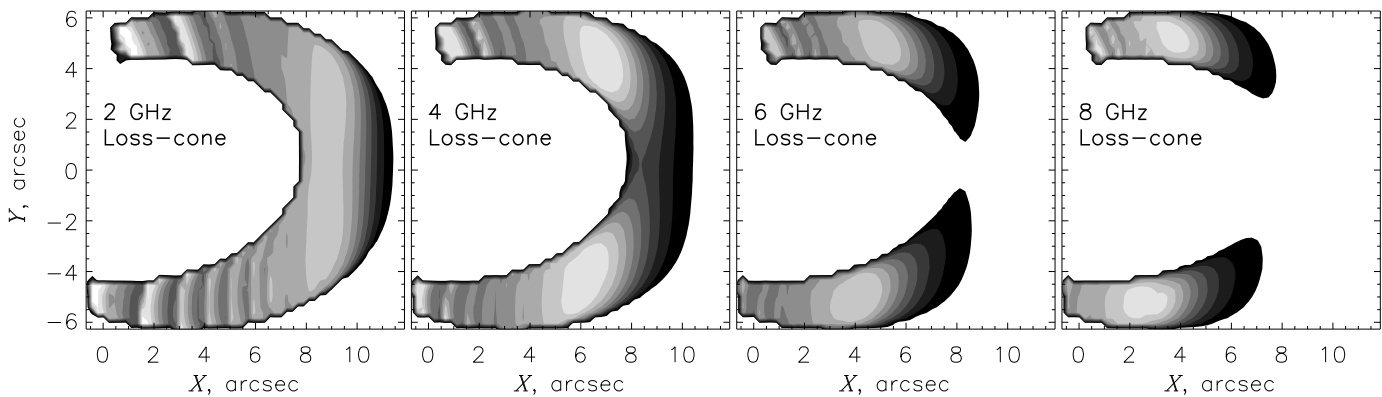}}
\caption{The same as in Fig. \protect\ref{Fig11}, for a loop located near the solar limb.}
\label{Fig12}
\end{figure*}


\subsection{Harmonic structure}\label{harm}
The GS emission from a homogeneous source can demonstrate an oscillatory spectral structure in the low-frequency range ($f/f_{\mathrm{B}}\lesssim 10$), when the emission intensity increases at a narrow spectral region near the harmonics (small integer multiples) of the electron cyclotron frequency \citep[see, e.g., the figures in the articles of][]{Ramaty_1969, ben92, FlMel_2003b, FlMel_2003a, Fl_Kuzn_2010}. In an inhomogeneous source, however, this harmonic structure can be hidden because of natural smoothing: the resonance giving rise to a gyroharmonics at a given location will vary with frequency due to the spatially dependent resonant condition in the spatially non-uniform  magnetic field. Thus, even if a spectrum from a single pixel contains harmonics they often disappear after integration over even a relatively small part of the source. This is why no harmonic structure is present in either footpoint or looptop spectra in Figures~\ref{Fig04}, \ref{Fig06}, \ref{Fig08}, and \ref{Fig10} other than a number of extremely large peaks in Figures~\ref{Fig06} and \ref{Fig10} for the loss-cone case provided that the conditions for the electron cyclotron maser (ECM) instability \citep[see, e.g.,][]{Stepanov_1978, Wu_Lee_1979, Holman_1980, Hewitt_etal_1982, dul82, Sharma_Vlahos_1984, Wu_1985, Aschw_1990, Fl_Meln_1998, Fl_Arzner_2000, LaBelle_Treumann_2002, Treumann_2006, Kuznetsov_2011} are locally fulfilled. We do not consider the coherent ECM emission here concentrating on the harmonic structure of the incoherent GS emission.

In the simplest case of a uniform GS source, the harmonic structure is more prominently pronounced for the source  viewed at a quasitransverse direction relative to the source magnetic field. Note, that for the magnetic model adopted here, the harmonic structure is only expected from the footpoint and leg regions, while not from the looptop where the magnetic field is too weak for the gyroharmonics to be produced at the considered parameter regime; thus, the limb location of the loop is most favorable to produce distinct gyroharmonics, which is confirmed by comparing Figures~\ref{Fig11}-\ref{Fig12}.

These figures show the brightness maps at low frequencies for two loops locations---at the disk center and limb. As expected, there is no apparent harmonic structure in the images seen at the disk center, Fig.~\ref{Fig11}. In contrast, for the loop located near the limb, the images contain a number of bright stripes highlighting the isolines of the magnetic field strength corresponding to the gyroresonance conditions at a given frequency. At  2 GHz, for example, these stripes are clearly visible from the footpoints up to about half of the loop height. If one were gradually increasing  the frequency, for which the image has been computed, each stripe would demonstrate an apparent down-motion because same harmonic number requires proportionally larger magnetic field for a higher frequency. Accordingly, at higher frequency images, the stripes are shifted toward the footpoints, and their contrast (or amplitude of the intensity oscillations) decreases.

Apparently, having these stripes detected in real observations would offer a highly efficient way of model-independent quantitative measurement of the coronal magnetic field (note that   averaging along the line of sight plays here only a minor role since the GS optical depth is large at the gyro harmonics so the layer contributing to the emission is narrow implying almost uniform magnetic field along the essential part of the integration). In case of rather compact loop considered here this task requires a radio array with the spatial resolution about 1" at the low frequencies (where scattering of radio waves by coronal density inhomogeneities can additionally fuze the images, \citet{Bastian_1994}, however), which is comparable with the anticipated capability of the Frequency Agile Solar Radiotelescope  \citep[FASR,][]{Gary_2003}; although for the instruments currently under development (e.g., USSRT and EOVSA) detecting  the harmonic stripes can only be expected from the biggest flaring loops and/or for the cases with a more uniform magnetic field. Such favorable cases, although untypical, are not unlikely: \citet{Staehli_etal_1987} reported a total power data on a microwave burst obtained with a few single frequencies that demonstrated prominent harmonic-like enhancements at two frequencies. Furthermore, the OVSA total power data reveal clear horizontal stripes at the dynamic spectra of some events, e.g., 2003 June 15 flare \citep{Fl_etal_2008}; see also \citep{ben92}, which might indicate the gyroharmonic contribution; however, such an interpretation remains ambiguous unless confirmed by direct imaging data.

\section{Discussion}
We have considered microwave emission produced by moderately anisotropic electron distributions populating (uniformly or non-uniformly) a non-uniform symmetric dipole magnetic loop. We emphasize that for the adopted loop geometry, the mirror ratio is more than 10, so in most of the loop volume (except the close vicinity of the footpoints) the loss-cone angle defined by the adiabatic invariant (Figure~\ref{Fig02}b) is small and, accordingly,  the angular distribution of the fast electrons is close to the isotropic one, which we here conventionally call a ``moderate anisotropy''. But even this moderate anisotropy affects noticeably both images and spectral characteristics of the emission.

In particular, for the uniform electron distribution along the loop the pitch angle anisotropy enhances the optically thin emission from the footpoint vicinity for the loop located at the limb, while suppresses it for the loop located at the disk center. In the case of non-uniform electron distributions due to accumulation at the loop top, the relative contribution of the electrons in the vicinity of the footpoints decreases and, so, the effect of the anisotropy becomes less pronounced. Nevertheless, in most of the cases (except looptop emission, where the electron angular distribution is almost isotropic) the effect of anisotropy on the images, spectrum, polarization, and spectral index can easily be recognized. We note that for larger thermal density than adopted for our restricted modeling, which is indeed often the case \citep{Bastian_etal_2007}, the looptop emission will be strongly suppressed by the Razin effect (which is specified by the $n_0/B$ ratio and becomes especially strong at $f< 20n_0/B$), so the emission from the footpoints and legs of the loop, where the electron anisotropy is stronger than at the looptop, will dominate the emission; thus the anisotropy effect will be even stronger than for the cases discussed above, see on-line supplement Album. 

An interesting feature of the low-frequency images is the presence of the ``gyro-stripes'', which are indicative of distinct gyroharmonics produced at  certain heights changing with frequency (since a few small integer multiples of the local gyrofrequency produce strong enhancement of the flux density). Having radio imaging spectroscopy data with the spatial resolution sufficient to resolve such gyro-stripes would offer a nice model-independent diagnostics of the coronal magnetic field during flares. In the near future the Expanded OVSA will be capable of such measurements at least for relatively large flaring sources.

An important issue is anisotropy/non-uni\-for\-mi\-ty effect on the total power radio spectrum. Since the total power spectrum is the result of emission integration over the entire source area, the result of this integration depends on how the distinct spatially resolved  contributions are weighted, which, in turn, depends on the anisotropy, inhomogeneity, and the viewing angle. We found that all the spectrum peak, peak flux, polarization, and the spectral index depend on both anisotropy and electron distribution inhomogeneity, although the effects are counter-directed. In particular, the radio spectral index changes noticeably with frequency and its behavior depends noticeably on the pitch-angle anisotropy, e.g., one can see that the maximal value of the total power spectral index (Figure~\ref{Fig04}, right bottom) for the anisotropic case exceeds by roughly 1 that for the isotropic case.

We compared the spectral indices with asymptotic relativistic (synchrotron) and approximate Dulk-Marsh values as the observed radio spectral indices are often used to evaluate the fast electron energy spectral indices. We found that neither synchrotron nor the Dulk-Marsh  value is a good approximation to the true spectral index: although the synchrotron index does represent the true index at high frequencies (higher than considered in our plots), the Dulk-Marsh index is less meaningful even though it sometimes coincides with the true one at a given single frequency. Therefore, we do not confirm the conclusion made by \citep{Simoes_Costa_2010} that the Dulk-Marsh index is a quantitatively good approximation to the spectral index of the spatially integrated spectrum, which might be an artifact specific to their adopted source model.

\section{Conclusion}
We have introduced a flexible simulation tool capable of fast computing 3D models of the microwave emission utilizing recently developed fast GS codes and presented an example of its use for the microwave emission modeling. Considering a highly restricted parameter space (symmetric dipole magnetic loop, moderate anisotropy, weak or no Razin-effect) we have analyzed images, spectra, and polarization of the model radio emission in the view of the available and future (more complete) radio data. In particular, we note that the high-frequency spectral index does not have a unique value for a given energy spectral index of radiating electrons; instead, it noticeably varies with the frequency, the viewing angle, the anisotropy, and the inhomogeneity and also different for various parts of the source and the entire source. This implies that the use of the radio spectral index for constraining the electron energy spectral index is not that straightforward; in place, the electron spectrum recovery must rely on the forward fitting of the entire radio datacube (ideally, including polarization) like in the example presented by \cite{Fl_etal_2009}.

\acknowledgements
This work was supported in part by NSF grants AGS-0961867, AST-0908344, and NASA grants NNX10AF27G and NNX11AB49G to New Jersey
Institute of Technology, and by the RFBR grants 09-02-00226, 09-02-00624, and 11-02-91175. A.K. thanks the Leverhulme Trust for financial support.

\bibliographystyle{apj}

\section*{Appendix: Magnetic Model}

\begin{figure*}
\epsscale{2}\plotone{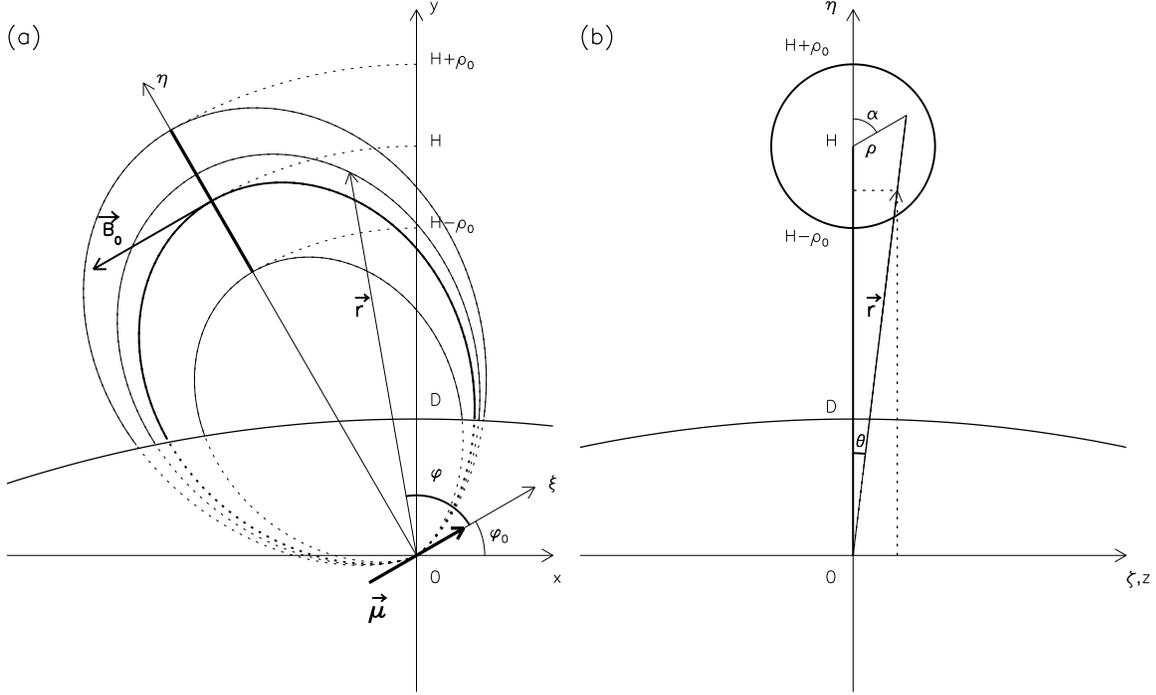}
\caption{\label{dipole_geometry}The magnetic dipole moment model implemented in the simulation tool. }
\end{figure*}

 Similar with other models used in the past \citep[e.g.][]{Antiochos_1976}, in our simulation tool the magnetic field is produced by magnetic dipole with a moment $\vect{\mu}$ that makes an adjustable angle $\pi/2-\varphi_0$ with its corresponding solar radius, and located below the solar surface at a depth $D$. The flaring loop is constructed around a central field line that is chosen to lie in the plane defined by the magnetic dipole vector and the local vertical, referred from now on as the central plane of the loop. As pictured in Figure \ref{dipole_geometry}, in the dipole's cartesian system of coordinates $\{\xi,\eta,\zeta\}$, which is defined as having the axis $\overrightarrow{O\xi}$ oriented along the magnetic dipole $\vect{\mu}$, and the axis $\overrightarrow{O\eta}$ in the central plane, the central magnetic field line may be parameterized as
\begin{eqnarray}
\label{centerline0}
\nonumber
\xi_c&=&H\sin^2\varphi\cos\varphi\\
\eta_c&=&H\sin^3\varphi,
\end{eqnarray}
where $\varphi\in[0,180]$ represents the angle between the position vector $\mathbf{r}_c=\{\xi,\eta,0\}$ and the dipole's axis, i.e. $\cos{\varphi}=\xi/r_c$, and $H$ is the height of the central line, i.e the maximum distance between the central line and the dipole's axis, which corresponds to $\varphi=\pi/2$.
In a solar coordinate system defined as having the axis $\overrightarrow{Oy}$ along the solar radius and the axis $\overrightarrow{Oz}$ identical with the axis $\overrightarrow{O\zeta}$, the same central line is expressed as
\begin{eqnarray}
\label{centerline}
\nonumber
x_c&=&H\sin^2\left(\varphi\right)\cos\left(\varphi+\varphi_0\right)\\
y_c&=&H\sin^2\left(\varphi\right)\sin\left(\varphi+\varphi_0\right).
\end{eqnarray}
Equations (\ref{centerline0}) and (\ref{centerline}) are a direct consequence of the fact that any magnetic dipole field line is described by the general polar equation $r=H\sin^2\varphi$ and that the dipole's system of coordinates is rotated relative to the solar system of coordinates by the angle $\varphi_0$ around the $\overrightarrow{Oz}$ axis, i.e.
\begin{eqnarray}
\label{transform}
\nonumber
x&=&\xi\cos{\varphi_0}-\eta\sin{\varphi_0}\\
y&=&\xi\sin{\varphi_0}+\eta\cos{\varphi_0}\\\nonumber
z&=&\zeta
\end{eqnarray}

The magnetic flux tube centered on the central field line is defined in terms of a circular cross section of radius $\rho_0$ that is normal to the central field at $\varphi=\pi/2$, i.e the magnetic loop-top. Hence, the cartesian coordinates of an arbitrary magnetic field line, $\mathbf{r}\equiv\{\xi,\eta,\zeta\}$, that intersects the loop-top cross section at distance $\rho$ from the central field line, may be expressed in terms of three convenient free parameters $\{\rho,\alpha,\varphi\}$ as
\begin{eqnarray}
\label{tubeline}
\nonumber
\xi&=&\sqrt{H^2 +2\rho H\cos{\alpha+ \rho^2}}\sin^2{\varphi}\cos{\varphi}\\
\eta&=&\left(H + \rho\cos{\alpha}\right)\sin^3{\varphi}\\\nonumber
\zeta&=&\rho\sin{\alpha}\sin^3{\varphi},
\end{eqnarray}
where, as shown in Figure \ref{dipole_geometry}b, $\alpha$ represents the fixed angle between the vector $\mathbf{\rho}$ and the dipole's vertical axis $\overrightarrow{O\eta}$, and $\varphi$ represents the variable angle between the vector $\mathbf{r}$ and the dipole moment $\vect{\mu}$ that is measured in the plane containing the field line, which makes the angle $\theta=\arctan\left(\zeta/\eta\right)=\rho\sin{\alpha}/\left(H+\rho\cos{\alpha}\right)$ with the central plane of the magnetic flux tube.

For $\rho=\rho_0$, the set of equations~(\ref{tubeline}) defines the envelope of the magnetic flux tube, which has a circular cross section shape only for $\varphi=\pi/2$, i.e. at the loop-top, while the shape of its cross section is continuously changing along the central field line.

The condition for an arbitrary point $\mathbf{r}$ to belong to a magnetic flux tube of height $H$ and top circular cross section of radius $\rho_0$, i.e. the condition for the magnetic field line passing trough the point $\{\xi,\eta,\zeta\}$ to intersect loop-top circular cross section, may be expressed as
\begin{eqnarray}
\label{include}
\left[H - \eta\left(1+ \frac{\xi^2 }{\zeta^2+\eta^2}\right)^{3/2}\right]^2+\zeta^2\left( 1+ \frac{\xi^2 }{\zeta^2+\eta^2}\right)^3\\\nonumber
\le\rho_0^2,
\end{eqnarray}
which has to be combined with the condition
\begin{equation}
\label{abovesurface}
x^2+(y-D+R_{Sun})^2+z^2 \ge R_{Sun}^2
\end{equation}
in order to determine if such flux tube point is also located above the solar surface.

Considering a normal cross section of the flux tube corresponding to an angle $\varphi$ measured in the central plane of the flux tube, the parameter
\begin{eqnarray}
\label{s}
s&=&\frac{1}{2}H\Big\{\cos\varphi \sqrt{1+3 \cos^2 \varphi}+\\\nonumber
&&\frac{1}{\sqrt{3}}\ln\left[\sqrt{3}\cos\varphi+\sqrt{1+3 \cos^2 \varphi}\right]\Big\},
\end{eqnarray}
may be used, instead of the angular coordinate $\varphi$, as a convenient flux tube longitudinal coordinate corresponding to the normal cross-section of interest.

The strength of the magnetic field is controlled by a unique adjustable parameter $B_0$ that defines the absolute value of the loop top magnetic field vector $\mathbf{B}_0$, which is perpendicular on the loop-top cross section. Since the vector $\mathbf{B}_0$ uniquely determines the dipole moment $\vect{\mu}=-\mathbf{B}_0H^3$, the magnetic field vector in any point characterized by the position vector $\mathbf{r}\equiv\{x,y,z\}$ relative to the dipole origin is given by
\begin{equation}
\mathbf{B}=\frac{3(\vect{\mu}\mathbf{r})\mathbf{r}-r^2\vect{\mu}}{r^5}.
\end{equation}

\end{document}